\documentclass[AMA,STIX1COL]{WileyNJD-v2}

\usepackage[titletoc,title]{appendix}
\usepackage{etoolbox}
\usepackage{bm}
\usepackage{amsmath}
\usepackage{fancyvrb} 
\usepackage{listings,color} 
\usepackage{moreverb}

\DeclareMathSymbol{\psi}{\mathalpha}{letters}{32}

\graphicspath{{image/}}

\usepackage[utf8]{inputenc}
\DeclareUnicodeCharacter{03C8}{\psi}

\newcommand\BibTeX{{\rmfamily B\kern-.05em \textsc{i\kern-.025em b}\kern-.08em
T\kern-.1667em\lower.7ex\hbox{E}\kern-.125emX}}

\articletype{Article Type}%

\received{<day> <Month>, <year>}
\revised{<day> <Month>, <year>}
\accepted{<day> <Month>, <year>}

\begin{document}

\title{A Full Bayesian Model to Handle Structural Ones and Missingness in Economic Evaluations from Individual-Level Data}

\author[1]{Andrea Gabrio*}
\author[2]{Alexina J. Mason}
\author[1]{Gianluca Baio}
\authormark{Andrea Gabrio \textsc{et al}}

\address[1]{\orgdiv{Department of Statistical Science}, \orgname{University College London}, \orgaddress{\state{London}, \country{UK}}}
\address[2]{\orgdiv{Department of Health Services Research and Policy}, \orgname{London School of Hygiene and Tropical Medicine}, \orgaddress{\state{London}, \country{UK}}}
\corres{*Andrea Gabrio, Corresponding address. \email{ucakgab@ucl.ac.uk}}
\presentaddress{Gower Street, London WC1E 6BT UK}

\abstract[Abstract]{
Economic evaluations from individual-level data are an important component of the process of technology appraisal, with a view to informing resource allocation decisions. A critical problem in these analyses is that both effectiveness and cost data typically present some complexity (e.g.~non normality, spikes and missingness) that should be addressed using appropriate methods. However, in routine analyses, simple standardised approaches are typically used, possibly leading to biased~inferences. 

We present a general Bayesian framework that can handle the complexity. We show the benefits of using our approach with a motivating example, the MenSS trial, for which there are spikes at one in the effectiveness and missingness in both outcomes. We contrast a set of increasingly complex models and perform sensitivity analysis to assess the robustness of the conclusions to a range of plausible missingness~assumptions. 

This paper highlights the importance of adopting a comprehensive modelling approach to economic evaluations and the strategic advantages of building these complex models within a Bayesian framework.
}


\jnlcitation{\cname{%
\author{Gabrio A.}, 
\author{Mason AJ.}, and
\author{Baio G.}
} (\cyear{2017}), 
\ctitle{A Full Bayesian Model to Handle Structural Ones and Missingness in Economic Evaluations from Individual-Level Data}, \cjournal{Statistics in Medicine}, \cvol{2017;00:1--6}.}

\maketitle


\section{Introduction}\label{intro}
Economic evaluation alongside Randomised Clinical Trials (RCTs) is an important and increasingly popular component of the process of technology appraisal \citep{NICE2013}. The typical analysis of individual level data involves the comparison of two interventions for which suitable measures of clinical benefits and costs are observed on each patient enrolled in the trial, often at different time points throughout the follow up. For simplicity, we generically term the clinical benefits as ``effectiveness'' and thus indicate the economic outcome variables as $(e,c)$.

Typically, effectiveness is measured through multi-attribute utility instruments (e.g.~the EQ-5D 3L: \url{http://www.euroqol.org}), the costs are obtained using clinic resource use records and both are summarised into cross-sectional quantities, e.g.~Quality Adjusted Life Years (QALYs). The main objective of the economic analysis is to combine the population average effectiveness and costs and use the precepts of decision theory to determine the most ``cost-effective'' intervention, given current evidence, as well as to assess the uncertainty in the decision-making process, induced by the uncertainty in the model inputs~\citep{Briggs2000,OHagan2004,Sculpher,Spiegelhalter2003,Claxtonb,Baioa,Jackson,Briggs2006,Spiegelhalterb}.

In routine analyses, trial-based Cost-Effectiveness Analyses (CEAs) are usually performed under a frequentist approach in which the two outcome variables $(e,c)$ are modelled independently. Baseline adjustments are often included in the model using simple regression analyses \citep{Manca,Hunter,Agency}. 
This, often implicitly, assumes normality for the underlying cost and effectiveness data, or at least that the sample size is large enough for the population means to be (approximately) normally distributed. In addition, almost invariably the relationship between the outcomes and the baseline characteristics is assumed to be linear.

There are several potential issues with this setting: firstly, the assumption of independence between costs and effectiveness is often questionable. While this is a recognised problem in the CEA literature, particularly under a Bayesian framework \citep{OHagan,Nixon,Baioa}, and although it may introduce bias in the statistical modelling and, \textit{a fortiori}, in the economic evaluation \citep{OHagan,Thompson}, appropriate methods to deal with correlation have historically found little application in routine analyses \citep{Briggsf}. 

Secondly, because both costs and effectiveness are usually characterised by a large degree of skewness, the assumption of normality is unlikely to hold and alternative approaches have been proposed in the literature. Examples include nonparametric bootstrapping \citep{Rascati} and, particularly within a Bayesian approach, the use of more appropriate parametric modelling \citep{Nixon,Thompson}. Nonparametric bootstrapping mostly relies on using simple averages that often give similar results to those assuming normality \citep{OHagan2003}. Conversely, modelling based on different parametric distributions (e.g.~Gamma for the costs and Beta for the QALYs) often allows improvement in the model fit to the observed data and appropriately captures skewness. 

Thirdly, data may exhibit spikes at one or both of the boundaries of the range for the underlying distribution. For example, some patients in a trial may not accrue any cost at all (i.e.~$c_i=0$), thus invalidating the assumptions for the Gamma distribution, which is defined on the range ($0,+\infty$). Similarly, we may observe individuals who are associated with perfect health, i.e.~unit QALY \citep{Basu}, which makes it difficult to use a Beta distribution, defined on the open interval $(0,1)$. A simple solution is to add/subtract a small constant $\epsilon$ to the entire set of observed values for the cost/effectiveness variable, thus artificially re-scaling it in the desired interval \citep{Cooper}. Despite being very easy to implement, this strategy is potentially problematic as the results are likely to be strongly affected by the actual choice of the scaling parameter $\epsilon$ and no clear guideline exists about the value to use (e.g.~$0.1, 0.01, \ldots$). In addition, when the proportion of these values is substantial, they may induce high skewness in the data and the application of simple methods may lead to biased inferences \citep{Mihaylova}. A more efficient solution suggested to handle this issue is the application of \textit{hurdle models} \citep{Ntzoufras,Mihaylova,Baio}. These are mixture models defined by two components: the first one is a mass distribution at the spike, while the second is a parametric model applied to the natural range of the relevant variable. Usually, a logistic regression is used to estimate the probability of incurring a ``structural'' value (e.g. 0 for the costs, or 1 for the QALYs); this is then used to weight the mean of the ``non-structural'' values estimated in the second component. Hurdle models have been discussed and applied in CEA mainly for handling structural zero costs \citep{Tooze,Harkanen,Baio}. 

Finally, individual level data from RCTs are almost invariably affected by the problem of missing data. Numerous methods are available for handling missingness in the wider statistical literature, each relying on specific assumptions whose validity must be assessed on a case-by-case basis. Whilst some guidelines exist for performing CEAs in the presence of missing outcome values \citep{ISPOR}, they tend not to be consistently followed in published studies \citep{Groenwold,Wood,Noble,Gabrio}. Analyses that are limited to the observed data (Complete Case Analysis, CCA) are inefficient and may yield biased inferences \citep{Briggsb,Mancab,Harkanen,Faria}. Multiple Imputation \citep[MI;][]{Rubina} is a more flexible method, which increasingly represents the \textit{de facto} standard in clinical studies \citep{DiazOrdaz,Burton}. In a nutshell, MI proceeds by replacing each missing data point with a value simulated from a suitable model. $M$ complete (i.e.~without missing data) replicates of the original dataset are thus created, each of which is then analysed separately using standard methods. The individual estimates are pooled using meta-analytic tools such as \textit{Rubin's rules} \citep{Rubina}, to reflect the inherent uncertainty in imputing the missing values. For historical reasons, as well as on the basis of theoretical considerations, the number of replicated datesets $M$ is usually in the range 5-10 \citep{Rubina,Schafer,Schaferbook}. 

As a consequence of the separation between the imputation and the analysis steps, MI requires the property of \textit{congeniality},~i.e. the imputation model needs to be specified as equally or less restrictive than the analysis model \citep{VanBuuren}. In addition, in many applications, MI is based upon assuming a \textit{Missing At Random} (MAR) mechanism,~i.e. the observed data can explain fully the reason for why some observations are missing. However, this may not be reasonable in practice (e.g.~for self-reported questionnaire data) and it is important to explore whether the resulting inferences are robust to a range of plausible \textit{Missing Not At Random} (MNAR) mechanisms, which cannot be explained fully by the observed data. Neither MAR nor MNAR assumptions can be tested using the available data alone and thus it is crucial to perform sensitivity analysis to explore how variations in assumptions about the missing values impact the results \citep{Carpenter, Molenberghsb}.

Building on the existent literature, we show how models that simultaneously account for different potential sources of bias can be efficiently specified under a full Bayesian framework, which has several advantages in comparison to a frequentist counterpart, specifically in health care technology assessments \citep{Spiegelhalterb, Baioa}. Firstly, by virtue of its modular nature, Bayesian modelling is very flexible, which means that a basic structure can be relatively easily extended to account for the increasing complexity required to formally allow for the several features described above. We exploit this in \protect\S\ref{models}. In addition, the Bayesian approach naturally allows for the principled incorporation of external evidence (e.g.~expert opinions) through the use of prior distributions. This is often crucial for conducting sensitivity analysis to a plausible range of missingness assumptions including MNAR \protect\citep{Daniels, Mason}, particularly when the evidence produced by the current study is limited, as is the case for small pilot trials, whose objective is to aid decision making about larger investments. Examples include the conduct of a full-scale trial, or the introduction in the market of a new cancer drug, based on the extrapolation of survival data produced over a short follow up.

Moreover, we note that MI can be considered as an approximation to a full Bayesian analysis on different levels. First, MI separates the imputation and analysis steps in two estimation procedures while, within a full Bayesian approach, the parameters of interest are estimated simultaneously with the imputation of the missing values and no additional analysis or \textit{ad hoc} pooling is necessary. Even though it has been shown that MI performs well in most standard situations, when the complexity of the analysis increases, a full Bayesian approach is likely to be a preferable option as it naturally allows the propagation of uncertainty to the wider economic model and to perform sensitivity~analysis. Second, due to the small number of replicates that are kept in practice, MI can be thought of as a fully Bayesian analysis based on a few simulations. Interestingly, the often-quoted objection to Bayesian modelling, i.e.~that it is too computationally intensive in comparison to simpler frequentist counterparts, is likely to dissolve in the presence of extremely complex models, which would require tailor-made routines for the optimisation of non-standard multivariate likelihood functions, thus effectively surrendering their computational advantage over intensive but efficient sampling methods such as Markov Chain Monte Carlo (MCMC).

The main contribution of this work is to provide a unified framework that allows jointly tackling the features in CEA discussed above. We use a real case study based on a small pilot trial as our motivating example. Starting from the original analysis, we progressively expand our basic model. We specifically focus on appropriately modelling spikes at the boundary and missingness, as they have substantial implications in terms of inferences and, crucially, cost-effectiveness~results. The paper is structured as follows: in \protect\S\ref{study} we present the case study and describe the data. \protect\S\ref{models} defines the general structure of the statistical model used in the analyses and how it can be tailored to deal with the different features affecting the data. Initially, for simplicity, we present each model under a complete case scenario that will then be extended to account for missingness. \protect\S\ref{Res} compares the results from three alternative models under both a complete cases and all cases scenario assuming MAR. The robustness of the results to alternative MNAR assumptions is then explored. \protect\S\ref{EE} summarises the inferences for each model from a decision-maker perspective and compares their implications in terms of cost-effectiveness. \protect\S\ref{discussion} discusses the proposed framework and suggests some improvements for future work. Finally, the Appendix includes additional material related to model assessment and the computer code for our analysis.

\section{Case Study: The MenSS Trial}\label{study}
We use as a motivating example the MenSS trial \citep{Hunterb}, a pilot RCT conducted in the UK public care sector to evaluate the cost-effectiveness of a new interactive digital intervention (the Men's Safer Sex website, MenSS). This new intervention provides individually tailored advice on barriers to condom use to reduce the incidence of Sexually Transmitted Infections (STIs) in young men. A total of 159 men aged 16 years and over with female sexual partners and recent unprotected sex or suspected acute STI were recruited from three English sexual health clinics. Participants were randomised to receive either usual clinical care only (comparator, $n_{1}=75$), or a combination of usual care and the MenSS website (active intervention, $n_2=84$). Sexual health related resource use was collected via participant responses to questionnaires at 3, 6 and 12 months. For each individual $i$, utility scores $u_{ij}$ were computed based on a generic health related quality of life questionnaire, the EQ-5D 3L, collected at baseline ($j=0$) and then at 3, 6 and 12 months ($j=1,2,J=3$). QALYs and total costs (measured in \pounds) were calculated by combining the utilities $u_{ij}$ and costs $c_{ij}$ collected at each time point as

\begin{equation}\label{QALY}
e_{i}=\sum_{j=1}^{J}(u_{ij}+u_{i\,j-1})\frac{\delta_{j}}{2} \;\;\; \text{and} \;\;\;\  c_{i}=\sum_{j=1}^{J}c_{ij},
\end{equation}
where $\delta_{j}=\frac{\text{Time}_{j}-\text{Time}_{j-1}}{\text{Unit of time}}$ is the fraction of the time unit (12 months) between consecutive measurements, e.g.~$\delta_2=(\mbox{6 months}-\mbox{3 months})/\mbox{12 months}=0.25$. For the utilities, this approach is often referred to as the \textit{Area Under the Curve} \citep[AUC;][]{Drummond}.

\begin{figure}[!h]
\centering
\includegraphics[scale=1.3]{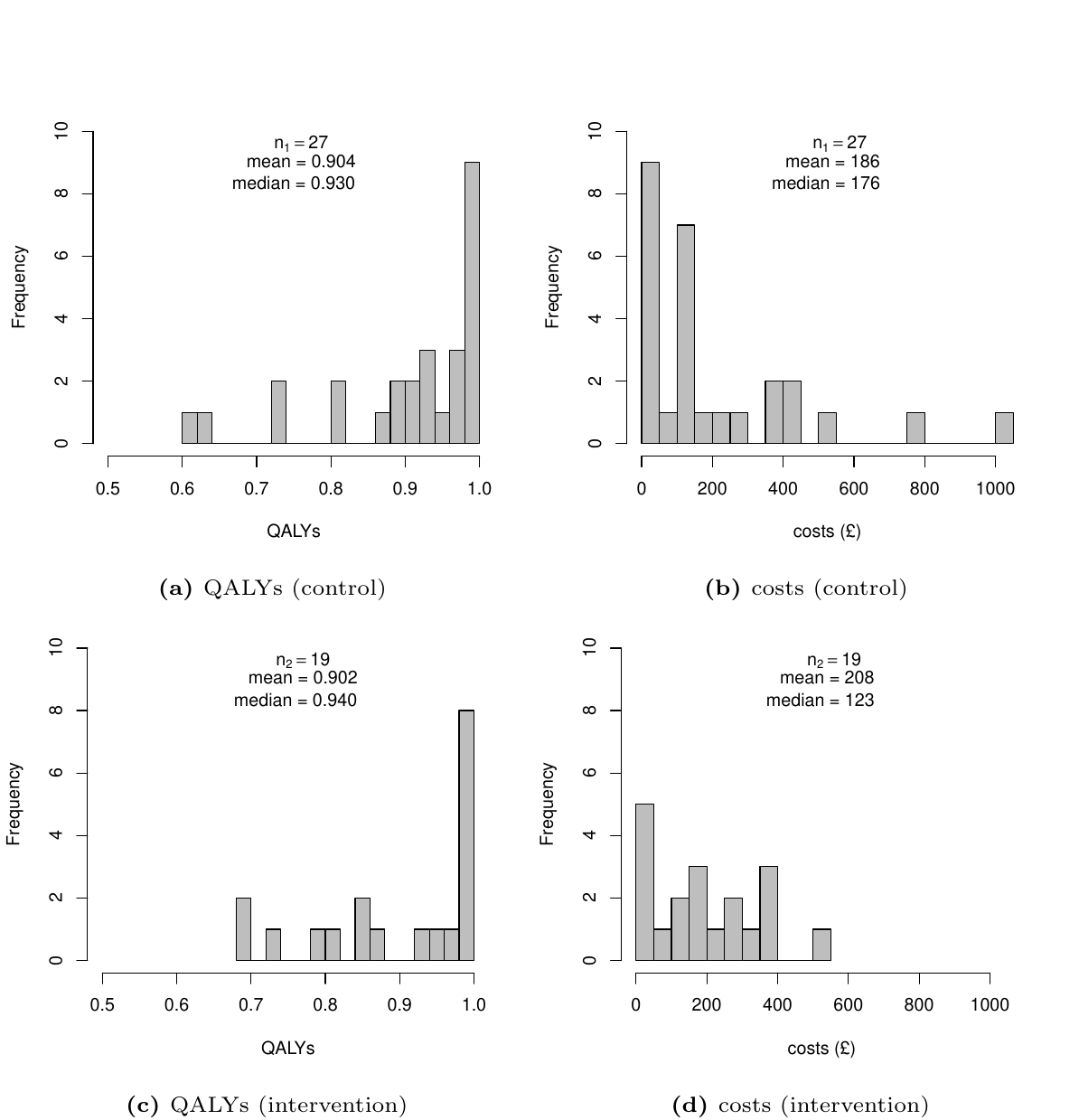}
\caption{Histograms of the distributions of the complete case QALYs and costs, expressed in \pounds{}, in the control (panels a-b) and intervention~(panels c-d) group. For both variables and in both arms, skewness of the observed data is apparent.}\label{hist}
\end{figure}

The number of participants completing utility and cost questionnaires at every time point was 27 (36\%) and 19 (23\%) for the control and intervention group, respectively. Figure~\ref{hist} shows the histograms of the distributions of the complete case QALYs and costs in the control (panels a-b) and intervention (panels c-d)~group, respectively.

\begin{center}
FIGURE 1 HERE
\end{center}

The data clearly present some of the features described in \protect\S\ref{intro}. A relatively high degree of skewness characterises the empirical distributions of QALYs and costs in both treatment groups. In particular, the substantial proportion of individuals incurring a perfect health status (which we term ``structural ones'') observed in both control (33\%) and intervention (42\%) groups effectively induces spikes at 1 in the QALYs. Finally, a large proportion of missingness characterises both cost and utility data due to poor follow-up rates. A statistical summary of the observed cases and missingness levels for the utility and cost variables by follow up period is shown in Table~\ref{Tpattern} (recall that baseline data are collected for the utilities~only). 

\begin{center}
TABLE 1 HERE
\end{center}

The original analysis was performed under a frequentist approach using standard OLS regression \protect\citep{Hunterb}. Baseline utility regression adjustment was incorporated in the model, assuming MAR for all variables and restricting the analysis to the complete cases. However, most of the features described in \protect\S\ref{intro} were not explicitly taken into account.

\section{Modelling Framework}\label{models}
In this section, we firstly present our general modelling framework for cost-effectiveness data. The model improves the typical approach used in routine analyses by accounting for correlation between the outcomes. Then it is extended to handle structural values and missing data using three alternative specifications with increasing complexity. Throughout, we refer to our motivating example to demonstrate the flexibility of our full Bayesian approach in dealing with the idiosyncrasies highlighted above; we also note that these are likely to be encountered in many practical cases, thus making our arguments applicable in general.

Assume that some patient-level data are collected from a trial on $i=1,\ldots,n$ individuals who are randomly allocated to either a control ($t=1$) or intervention ($t=2$) group, with sample sizes $n_1$ and $n_2$, respectively. We denote by $e_{it}$ and $c_{it}$ the effectiveness and cost outcome variables for the $i-$th person in group $t$ of the trial. To simplify the notation, unless necessary, we suppress the treatment subscript~$t$. 

To account for correlation between the outcomes, in general we can specify the joint distribution $p(e,c)$ as:
\begin{equation}
p(e,c)=p(c)p(e\mid c)=p(e)p(c\mid e),\label{factorisation}
\end{equation} 
where, for example, $p(e)$ is the \textit{marginal} distribution of the effectiveness and $p(c\mid e)$ is the \textit{conditional} distribution of the costs given the effectiveness \citep{Nixon}. Note that while it is possible to use interchangeably either factorisation in Equation~\ref{factorisation}, without loss of generality, we describe our analysis in the following through a marginal distribution for the effectiveness (QALYs) and a conditional distribution for the costs. 

For each individual we consider a marginal distribution $p(e_i\mid\bm\theta_e)$ indexed by a set of parameters $\bm\theta_e$ comprising a \textit{location} $\phi_{ie}$ and a set of \textit{ancillary} parameters $\bm\psi_e$ typically including some measure of \textit{marginal} variance, $\sigma^2_e$. We can model the location parameter using a generalised linear structure, e.g.
\[ g_e(\phi_{ie})= \alpha_0 \,\,[+ \ldots], \]
where $\alpha_0$ is the intercept and the notation $[+\ldots]$ indicates that other terms (e.g.~quantifying the effect of relevant covariates) may or may not be included in the model. In the absence of covariates or assuming that a centered version $x_i^*=(x_i-\bar{x})$ is used, the parameter $\mu_e=g_e^{-1}(\alpha_0)$ represents the population average effectiveness.

For the costs, we consider a conditional model $p(c_i\mid e_i,\bm\theta_c)$, which explicitly depends on the effectiveness variable, as well as on a set of quantities $\bm\theta_c$, again comprising a location and ancillary parameters. Note that in this case $\bm\psi_c$ includes a \textit{conditional} variance $\tau^2_c$, which can be typically expressed as a function of the marginal variance $\sigma^2_c$  \citep{Nixon, Baioa}. The location can be modelled as a function of the effectiveness variable as:
\[g_c(\phi_{ic})=\beta_{0}+\beta_{1}(e_{i}-\mu_{e})\,\,[+\ldots]. \]
Here, $(e_i-\mu_e)$ is the centered version of the effectiveness variable, while $\beta_{1}$ quantifies the correlation between costs and effectiveness. Assuming other covariates are either also centered or absent, $\mu_c=g_c^{-1}(\beta_{0})$ is the population average~cost.

\begin{figure}[!h]
\vspace*{-1cm}
\centering
\includegraphics[scale=1.5]{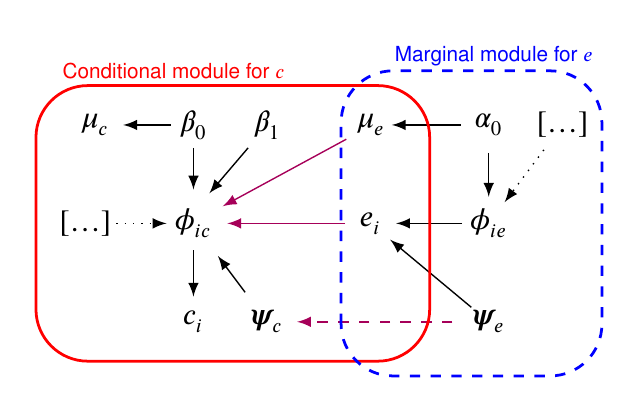}
\caption{Joint distribution $p(e,c)$, expressed in terms of a marginal distribution for the effectiveness and a conditional distribution for the costs, respectively indicated with a solid red line and a dashed blue line. The parameters indexing the corresponding distributions or ``modules'' are indicated with different Greek letters, while $i$ denotes the individual index. The solid black and magenta arrows show the dependence relationships between the parameters within and between the two models, respectively. The dashed magenta arrow indicates that the ancillary parameters of the cost model may be expressed as a function of the corresponding effectiveness parameters. The dots enclosed in the square brackets indicate the potential inclusion of other covariates at the mean level for both modules.}\label{model}
\end{figure}

Figure~\ref{model} shows a graphical representation of the general modelling framework described above. The effectiveness and cost distributions are represented in terms of combined ``modules'' --- the blue and the red boxes --- in which the random quantities are linked through logical relationships. This ensures the full characterisation of the uncertainty for each variable in the model. Notably, this is general enough to be extended to any suitable distributional assumption, as well as to handle covariates in either or both the modules.

\begin{center}
FIGURE 2 HERE
\end{center}

In the rest of the section, we present three alternative specifications of the general structure depicted in Figure~\ref{model} to model effectiveness and cost data. These are 1) Normal marginal for the effectiveness and Normal conditional for the costs (which is identical to a Bivariate Normal distribution for the two outcomes); 2) Beta marginal for the effectiveness and Gamma conditional for the costs; and 3) Hurdle Model. First, we present each assuming a complete cases scenario and then extend the structure to all cases (complete and partially observed), considering both MAR (for all models) or alternative MNAR scenarios (for the Hurdle Model only).

\subsection{Complete Cases}\label{foc}
\subsubsection{Bivariate Normal}\label{BN}
\protect
Arguably, the easiest way of jointly modelling two variables is to assume Bivariate normality, which in our context can be factorised into marginal and conditional Normal distributions for $e_i$ and $c_i\mid e_i$. This is the closest modelling structure to those underpinning a typical frequentist analysis, while also accounting for potential correlation between the outcomes. 

In line with current recommendations (and the original analysis of the MenSS trial), we adjust for the baseline utilities --- using a centered version~($u_{i0}-\bar{u}_{0}$). We model $e_i\mid\bm\theta_e \sim\mbox{Normal}(\phi_{ie},\sigma^2_e)$, using an identity link function for the location~parameter
\[ g_e(\phi_{ie})=\phi_{ie} =  \alpha_0 +\alpha_1(u_{i0}-\bar{u}_{0}).\] 
Here, the parameter $\alpha_1$ quantifies the impact of the centered baseline utilities on the QALYs, while $\mu_e=\alpha_0$ and $\sigma^2_e$ represent the marginal (population level) mean and variance, respectively.

As for the costs, we model $c_i \mid e_i,\bm\theta_c \sim \mbox{Normal}(\phi_{ic},\tau^2_c)$, where the conditional mean and variance are defined as
\[  g_c(\phi_{ic})=\phi_{ic}=\beta_0+\beta_1(e_{i}-\mu_{e}) \qquad \mbox{ and } \qquad \tau^2_c=\sigma^{2}_{c}-\sigma^{2}_{e}\beta_1^2. \] 
The model parameters are thus $\bm\theta_e=(\alpha_0,\alpha_1,\sigma^2_e)$ and $\bm\theta_c=(\beta_0,\beta_1,\mu_e,\sigma_c^2,\sigma^2_e)$ --- note that the marginal mean and variance of the effectiveness link the two modules and therefore feature in both sets of parameters. 

The model is completed by assigning suitable prior distributions to the elements of $\bm\theta=(\bm\theta_e,\bm\theta_c)$; for example, independent Normal priors can be assumed for the regression parameters, while Uniform or Half-Cauchy priors can be assigned on the scale of the standard deviations~\citep{Gelmanp}.

\subsubsection{Beta-Gamma}\label{BG}
\protect
The second model we consider assumes a Beta marginal for the QALYs and a Gamma conditional for the costs. In particular, we parameterise the Beta distribution in terms of the mean $\phi_{ie}$ and the scale parameter $\tau_{ie}=\left(\frac{\phi_{ie}(1-\phi_{ie})}{\sigma^2_e}-1\right)$ as $e_i\mid\bm\theta_e\sim\mbox{Beta}\left(\phi_{ie}\tau_{ie}, (1-\phi_{ie})\tau_{ie}\right)$ and model the location as
\[ g_e(\phi_{ie}) = \mbox{logit}(\phi_{ie})=\alpha_0+\alpha_1(u_{i0}-\bar{u}_{0}).  \]

The costs are modelled as $c_i\mid e_i,\bm\theta_c \sim \mbox{Gamma}\left(\phi_{ic}\tau_{ic}, \tau_{ic}\right)$, where the shape parameter is defined as the product of the location $\phi_{ic}$ and the rate $\tau_{ic}$. The generalised linear model for the location is 
\[ g_c(\phi_{ic})=\log(\phi_{ic})=\beta_0+\beta_1(e_{i}-\mu_{e}). \]

The marginal means for the QALYs and total costs can then be obtained using the respective inverse link functions
\[ \mu_e=\frac{\text{exp}(\alpha_0)}{1+\text{exp}(\alpha_0)}  \qquad \mbox{ and } \qquad   \mu_c=\text{exp}(\beta_0).  \]
Notice that, in comparison to the Bivariate Normal of \S\ref{BN}, the Beta-Gamma model reflects more closely the range and skewness of the observed data. Nevertheless, this modelling structure also fails to directly account for the structural values, e.g.~unit QALYs. In the presence of structural values, it is necessary to rescale the observed data, e.g.~by applying the Beta model to $e_i^*=e_i-\epsilon$ for some $\epsilon\rightarrow 0$.

The model is again completed by placing suitable priors on the parameters. Typically, it is easier to encode genuine prior information on the natural scale of the parameters \citep{Baio,BCEABook}. For example, we can use independent Normal priors for the regression coefficients $(\alpha_1,\beta_1)$, Uniform priors on suitable scales for $(\mu_c,\mu_e)$ and a Uniform or Half-Cauchy prior for $\sigma_c$. Notice, however, that a little more care is needed in defining a prior distribution for $\sigma_e$. In fact, by the mathematical properties of the Beta distribution, the variance is bounded by a function of the mean
\[ \sigma^2_e \leq\mu_e(1-\mu_e)=\upsilon. \]
Consequently, we can place an informative prior on the standard deviation $\sigma_{e}\sim\mbox{Uniform}(0,\sqrt{\upsilon})$, which coupled with a prior for $\mu_e$ induces a suitable prior for $\tau_e$ as well. Note that even by starting with vague distributions for $(\mu_e,\sigma_e)$, the resulting prior for the Beta scale $\tau_e$ may not be vague at all. 

\subsubsection{Hurdle Model}\label{HU}
To overcome the limitations of the model in \S\ref{BG} in terms of the structural ones, we expand it to a hurdle version. Specifically, for each subject in the trial $i=1,\ldots,n$ we define an indicator variable $d_{ie}$ taking value 1 if the $i-$th individual is associated with a unit QALYs ($e_{i}=1$) and $0$ otherwise ($e_{i}<1$). This is then modelled as
\begin{align}
d_{ie}:=\mathbb{I}(e_{i}=1)&\sim\mbox{Bernoulli}(\pi_{ie}) \nonumber \\
\mbox{logit}(\pi_{ie})&= \gamma_0 + \gamma_1(u_{i0}-\bar{u}_{0})\,\,[+\ldots],\label{linpred_hurdle}
\end{align}
where $\pi_{ie}$ is the individual probability of unit QALYs, which is estimated on the logit scale as a function of a baseline parameter $\gamma_0$ and the centred baseline utilities, whose effect is captured by the parameter $\gamma_1$. As for the effectiveness and cost models, other covariates can be additively included in the model of $d_{ie}$. We specifically distinguish the baseline utilities from any other covariate as they are likely to be particularly informative in predicting whether an individual is associated with a structural one in the QALYs. All the logistic regression parameters should be given suitable prior probability distributions (e.g.~Normal). Within this framework, the quantity
\begin{equation}
\bar\pi_e=\frac{\mbox{exp}(\gamma_0)}{1+\mbox{exp}(\gamma_0)}\label{weight}
\end{equation}
represents the estimated marginal probability of unit QALYs. The parameters $\bar\pi_e$ and $(1-\bar\pi_e)$ in effect represent the weights used to mix the two components. 

Depending on the value of $d_{ie}$, we can partition the observed data on the QALYs into two subsets. In the first subset, defined as the $n^1$ subjects for whom $d_{ie}=1$, we define a variable $e^1_i=1$. Conversely, the second subset consists of the $n^{<1}=(n-n^1)$ subjects for whom $d_{ie}=0$ and for these individuals we define a variable $e^{<1}_i$. Because this is less than 1, we can model it directly using a Beta distribution characterised by an overall mean $\mu_e^{<1}$, in line with the specification we have shown in \S\ref{BG}. Using the estimated value for $\bar\pi_e$ from Equation~\ref{weight}, we can compute the overall population average effectiveness measure in both treatment groups $\mu_{et}$ as the linear combination
\begin{equation*}
\mu_{et}=(1-\bar\pi_{et})\mu^{<1}_{et}+\bar\pi_{et}. 
\end{equation*}
In the absence of structural zeros, the conditional model for the costs is exactly as specified in \S\ref{BG}.

\subsection{All Cases}\label{partobs}
When missingness occurs in the QALYs and cost variables, no change to the model structure is required under MAR for both the Bivariate Normal and Beta-Gamma specifications. For the Hurdle Model, when $e_i$ is missing, it is not possible to directly define the value for $d_{ie}$. However, unit QALYs can only be observed if $u_{ij}=1$ for all time points $j=0,\ldots,J$. Consequently, if an individual $i$ is such that $u_{ij}$ is missing at some time point $j$ and $u_{ij}\neq 1$ at any other time point, then by necessity $d_{ie}=0$. However, for all individuals having $u_{ij}=1$ at all observed time points but with at least one missing value at some other time point, then $d_{ie}$ is unknown.

Incomplete covariates need to be explicitly modelled to impute their missing values. For simplicity, we consider the case where the only covariate included in the model is the baseline utility; however, the same approach can be extended to any other type of partially-observed covariates. In the Bivariate Normal and the Beta-Gamma formulations, we can handle missingness in $u_{i0}$ by assuming a suitable model. One simple choice is to assume the same distribution for $u_{i0}$ as for the outcome $e_i$, i.e.~Normal or Beta, respectively --- for example, Appendix \ref{code} shows the implementation for the Beta-Gamma model.

Similarly to \S\ref{HU}, we can formulate another hurdle model for $u_{i0}$. More specifically, first we specify a model for the individuals with a non-unit utility value. Again, a simple solution is to base this on the same distributions assumed for the QALYs. Secondly, we estimate the probability of observing a structural one in the utilities as
\begin{align*}
d_{iu}:=\mathbb{I}(u_{i}=1)&\sim\mbox{Bernoulli}(\pi_{iu}) \nonumber \\
\mbox{logit}(\pi_{iu})&= \eta_0\,\,[+\ldots]
\end{align*}
where $d_{iu}$ is the indicator variable for the structural ones in the baseline utilities.

\subsubsection{Sensitivity analysis (MNAR)}\label{mnar}
Finally, Hurdle Models also offer a convenient framework for exploring the robustness of the results to some departures from MAR and therefore allow to perform a simple type of sensitivity analysis to the missingness assumptions. Two relevant cases are: 
\begin{itemize}
\item[\textit{a)}] the individuals for whom utility values are missing throughout the follow up, i.e.~$u_{ij}=\mbox{\texttt{NA}}$ for all $j=1,\ldots,J$;
\item[\textit{b)}] the individuals for whom all the observed utilities are equal to 1, but with at least one time point $j$ at which $u_{ij}=\mbox{\texttt{NA}}$.
\end{itemize}
For both these cases, it is impossible to compute the value of the indicator $d_{ie}$ according to the information from the observed data. However, we can arbitrarily set the value of $d_{ie}$ to either 1 or 0 using different configurations, e.g.~by varying the number of structural values potentially observed in a given scenario. Since these configurations are based on assumptions about the missing values that cannot be verified from the data at hand (but are in fact arbitrarily set by the experimenter), they effectively represent a way to assess the robustness of the results to some departures from MAR.

In the MenSS trial, there are $n^{*}_{1}=13 \; (12\%)$ individuals in the control and $n^{*}_{2}=22 \; (26\%)$ in the intervention group who fall within case \textit{a} or \textit{b}. Thus, we perform sensitivity analysis by defining a set of alternative MNAR assumption scenarios for these individuals and assess the robustness of the results across them. The four different scenarios considered are summarised in Table~\ref{tmnar}:

\begin{center}
TABLE 2 HERE
\end{center}
We choose these scenarios in order to assess how different ``extreme''  combinations of the number of potential structural ones in the intervention and control group can impact the results.

\section{Results}\label{Res}
We fitted all models using \texttt{JAGS}, \citep{Plummer}, a software specifically designed for the analysis of Bayesian models using Markov Chain Monte Carlo (MCMC) simulation, which can be interfaced with \texttt{R} through the package \texttt{R2jags} \citep{Su}. Samples from the posterior distribution of the parameters of interest generated by \texttt{JAGS} and saved to the \texttt{R} workspace are then used to produce summary statistics and plots. We ran two chains with 20,000 iterations per chain, using a burn-in of 10,000, for a total sample of 20,000 iterations for posterior inference. For each unknown quantity in the model, we assessed convergence and autocorrelation of the MCMC simulations using diagnostic measures such as the \textit{potential scale reduction factor} and the \textit{effective sample size} \citep{Gelman2}. The total running time required for the models to produce representative samples from the posterior distributions of interest ranged from 5 to 10 minutes.

Alternative prior distributions were considered to assess the sensitivity of the inferences to different vague prior specifications (e.g.~both Uniform and Half-Cauchy distributions for the standard deviations or different values for the variance of normally-distributed regression parameters). Results were robust to these specifications. Although the Hurdle Model as described in~\S\ref{HU} cannot be directly written in \texttt{JAGS}, it can be implemented using a simple ``coding trick''. Appendix~\ref{code} presents all the technical details and the \texttt{JAGS} script.

\subsection{Complete and All Cases (MAR)}\label{MAR}
Following the original analysis, we first consider only the complete cases and adjust for baseline utilities at the mean level for the QALYs in each model. For the Hurdle Model, in addition to the centered baseline utility, we include in the linear predictor of Equation~\ref{linpred_hurdle} three more categorical covariates (age, ethnicity and employment status). These are used to estimate the probability of structural ones in the QALYs. We then extend the framework to all cases under MAR, where the baseline utilities are explicitly modelled as detailed in~\S\ref{partobs} and again as functions of age, ethnicity and employment~status. 

\begin{figure}[!h]
\centering
\includegraphics[scale=1.5]{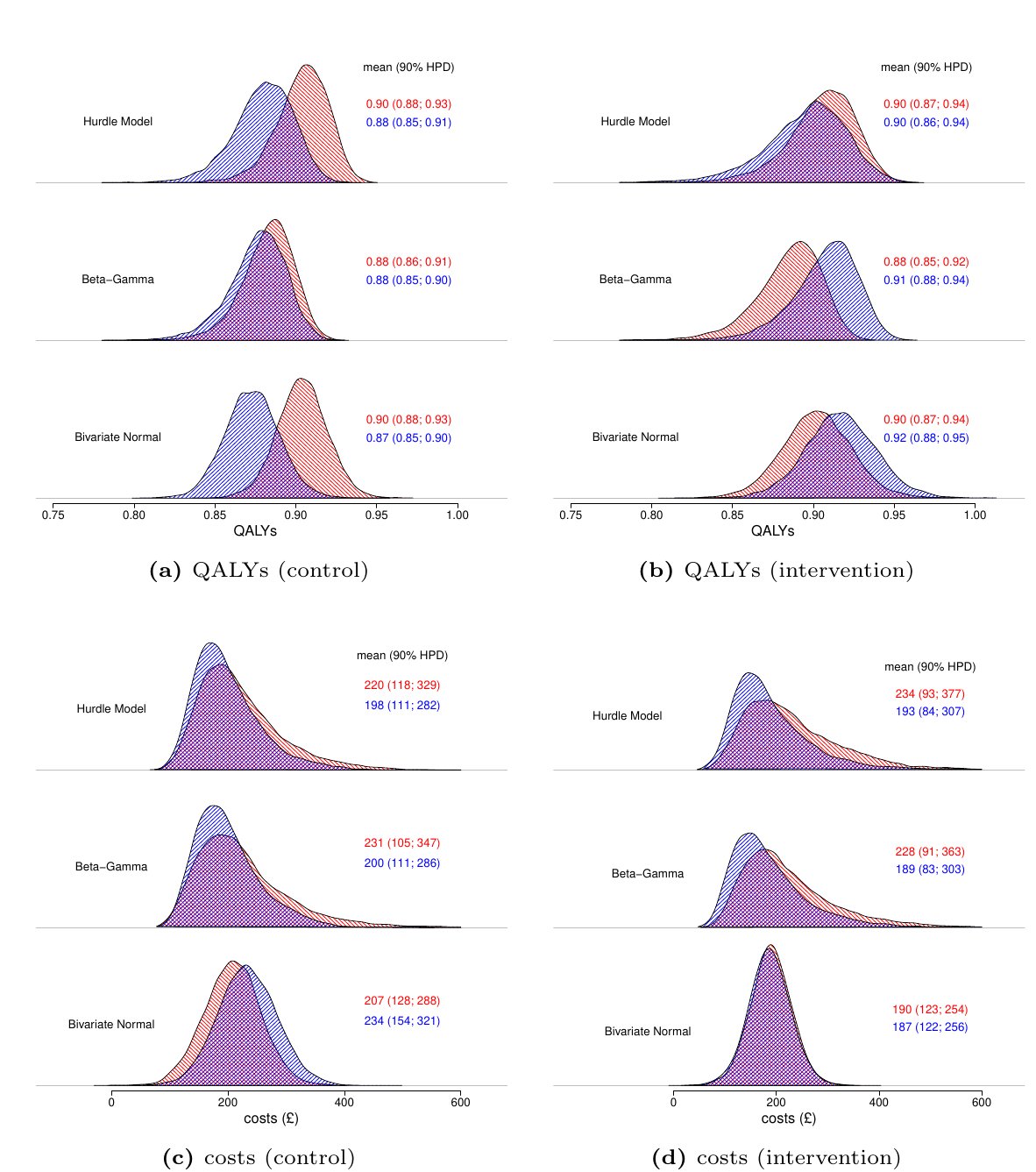}
\caption{Posterior distributions for the marginal mean parameters of the QALYs (panels a-b) and cost variables (panels c-d), expressed in \pounds{}, in each group of the trial under either a complete (red) and all (blue) cases scenario. The posterior results are presented for all model specifications considered (Bivariate Normal, Beta-Gamma and Hurdle Model) and for each of these the posterior mean estimates and associated 90\% Highest Posterior Density (HPD) interval bounds are reported.}\label{means}
\end{figure}

Figure~\ref{means} shows the posterior distributions of the mean QALYs and costs for both treatment groups under a complete (red) and all (blue) cases scenario for each model, assuming MAR.

\begin{center}
FIGURE 3 HERE
\end{center}

The posterior distributions of the mean QALYs (panels a-b) present some discrepancies between the complete and all cases scenarios, with magnitude varying according to the treatment group and model considered. In general, the results for all cases are lower in the control group and higher in the intervention group in comparison to those obtained using the complete cases. As for the mean costs (panels c-d), the results associated with a Gamma distribution are substantially more skewed compared to those obtaining using the Normal model, especially in the intervention group. In addition, the Gamma model typically leads to mean estimates that are systematically lower under the all cases scenario.

We compare the fit of the different models using the Deviance Information Criterion \citep[DIC;][]{Spiegelhalter}. The DIC is a measure of comparative predictive ability based on the model deviance and a penalty for model complexity. When comparing a set of models based on the same data, the one associated with the lowest DIC is the best-performing, among those assessed. There are different ways of constructing the DIC in the presence of missing data, which means that its use and interpretation are not straightforward \citep{Celeux,Masonb}. In our analysis, we consider a DIC based on the observed data and calculated only for the modules that are in common between the models, i.e.~excluding the contribution from the structural indicators for the Hurdle Model. The Bivariate Normal model is always associated with the highest DIC under both a complete and all cases scenarios ($536$ and $445$). The Beta-Gamma ($386$ and $60$) and, especially, the Hurdle model ($-50$ and $-2419$) substantially improve the model fit to the observed data.

\subsubsection{Imputations under MAR}
\begin{figure}[!h]
\centering
\hspace*{-1cm}
\includegraphics[width=18cm,height=18cm]{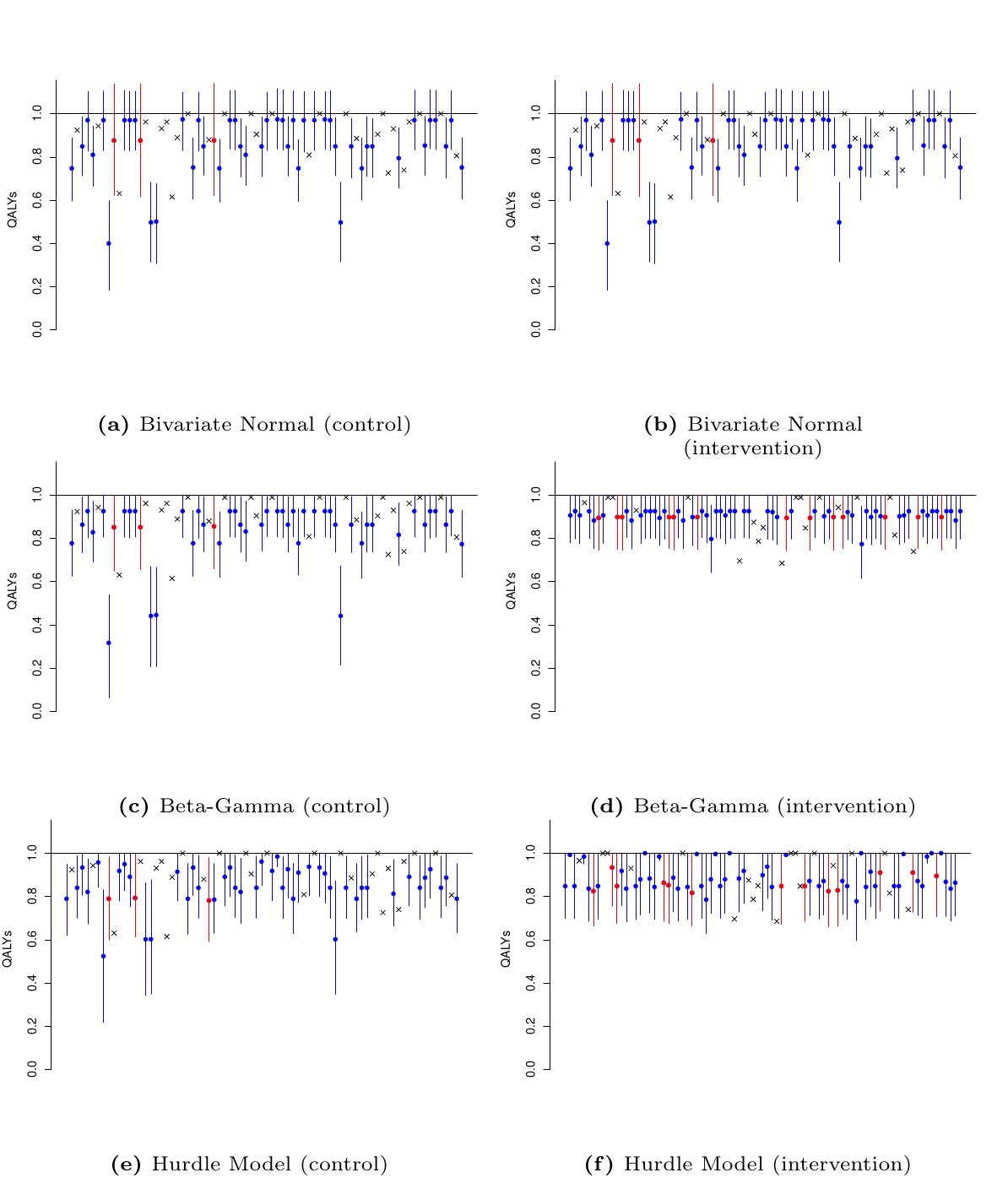}
\caption{Imputed QALYs in the control and intervention groups based on the Bivariate Normal, Beta-Gamma and Hurdle Model. Imputations are summarised in terms of posterior means and 90\% HPD intervals (coloured dots and lines) while an x symbol is used to denote the observed cases. Imputed values are also distinguished according to whether the corresponding baseline utilities were either observed (blue) or missing (red). The solid black line represents the upper bound for the utilities, set at the value of 1.}\label{imputed}
\end{figure}

Figure~\ref{imputed} depicts the observed QALYs in both treatment groups (indicated with black crosses) as well as summaries of the posterior distributions for the imputed values, obtained from each model. Imputations are distinguished based on whether the corresponding baseline utility value is observed or missing (blue or red lines and dots, respectively) and are summarised in terms of posterior mean and 90\%~Highest Posterior Density (HPD) intervals.

\begin{center}
FIGURE 4 HERE
\end{center}

There are clear differences in the imputed values and corresponding credible intervals between the three models in both treatment groups. Neither the Bivariate Normal nor the Beta-Gamma models  produce imputed values that capture the structural one component in the data. In addition, as to be expected, the Bivariate Normal fails to respect the natural support for the observed QALYs, with many of the imputations exceeding the unit threshold bound. These unrealistic imputed values highlight the inadequacy of the Normal distribution for the data and may lead to distorted inferences. Conversely, imputations under the Hurdle Model are more realistic, as they can replicate values in the whole range of the observed data, including the structural~ones. Imputed unit QALYs with no discernible interval are only observed in the intervention group due to the original data composition, i.e.~individuals associated with a unit baseline utility and missing QALYs are almost exclusively present in the intervention~group. 

\subsection{Sensitivity Analysis (MNAR)}\label{MNAR}
\begin{figure}[!h]
\centering
\hspace*{-0.5cm}
\includegraphics[scale=1.6]{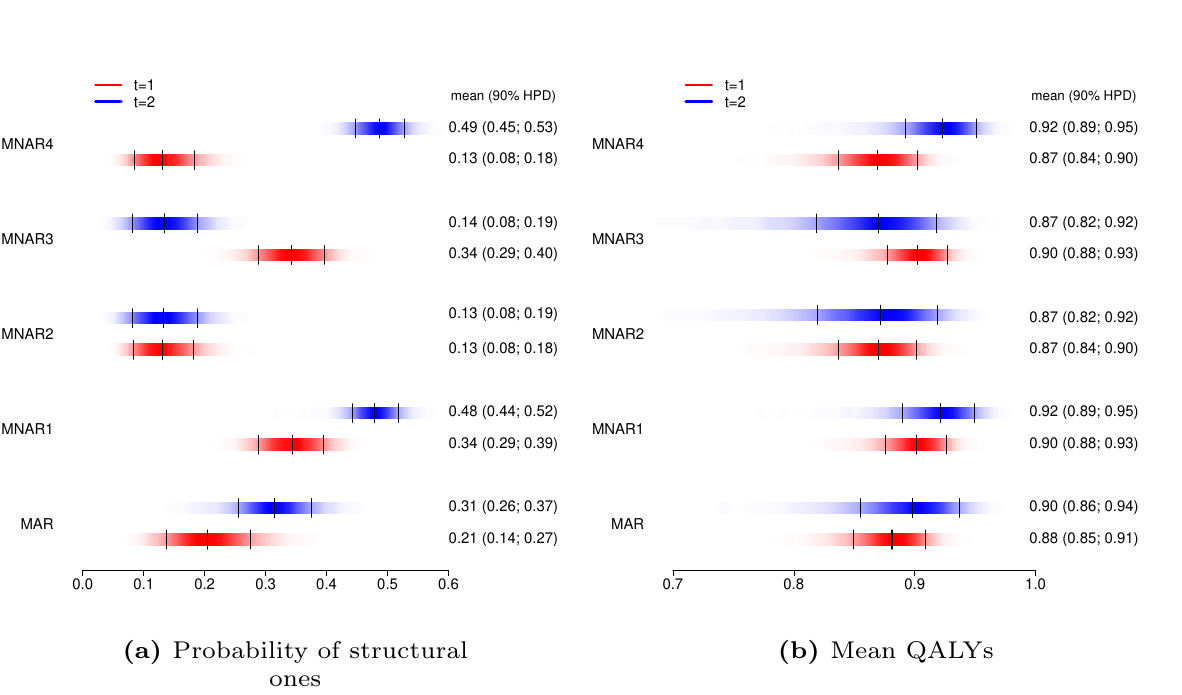}
\caption{Density strip plots for the posterior distributions of the probability of structural ones (panel a) and the marginal mean QALYs (panel b) under MAR and four alternative MNAR scenarios. For each scenario, results are presented for the control (red) and the intervention (blue) groups. Mean posterior values and associated 90\% HPD interval bounds are indicated with tick marks and reported aside for each quantity.}\label{mnar_plot}
\end{figure}

For each of the alternative MNAR scenarios described in~\S\ref{mnar}, as well as for the analysis under MAR,~Figure~\ref{mnar_plot} shows posterior density strips for the structural one probability $\bar\pi_e$ and the marginal mean QALYs $\mu_{e}$, in the control (red) and intervention (blue)~groups.

\begin{center}
FIGURE 5 HERE
\end{center}

Estimates under MAR indicate that the new intervention is associated with a probability of observing a structural one and a mean QALYs that are on average higher compared to the control. Although similar results are obtained under MNAR1, the estimated quantities are highly unstable across the other three MNAR scenarios.  Specifically, under MNAR2 the probability of structural ones is substantially reduced in both groups and induces a zero mean difference in the QALYs. Under MNAR3 and MNAR4 the differences between the estimated probabilities and mean QALYs in the two groups are increased in magnitude and lead to opposite mean differentials.

\section{Economic Evaluation}\label{EE}
We complete the analysis by assessing the cost-effectiveness of the new intervention with respect to the control, comparing the results of the different models under MAR (\S~\ref{MAR}) and the alternative MNAR scenarios explored for the Hurdle Model (\S~\ref{MNAR}). We specifically rely on the examination of the Cost-Effectiveness Plane \citep[CEP;][]{Black} and the Cost-Effectiveness Acceptability Curve \citep[CEAC;][]{VanHout} to summarise the economic analysis. 

\begin{center}
FIGURE 6 HERE
\end{center}

\begin{figure}[!h]
\centering
\hspace*{-0.5cm}
\includegraphics[scale=1.6]{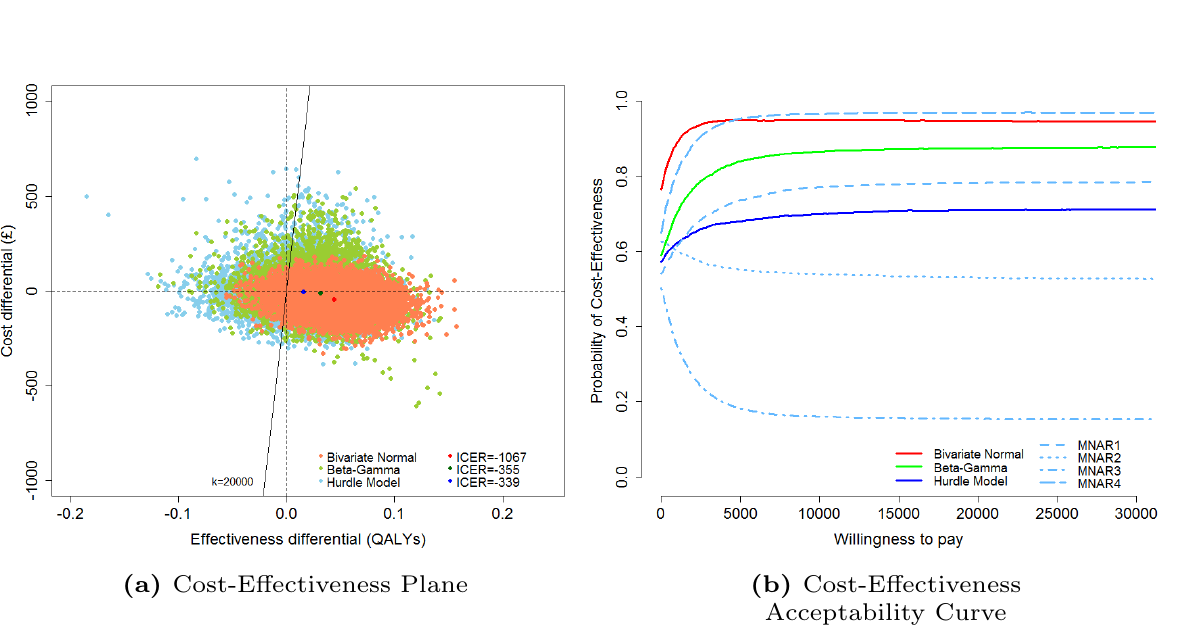}
\caption{CEPs (panel a) and CEACs (panel b) associated with the Hurdle (blue dots and line), Bivariate Normal (red dots and line) and Beta-Gamma (green dots and line) models. In the CEPs, the ICERs based on the results from the three model specifications under MAR are indicated with corresponding darker coloured dots, while the portion of the plane on the right-hand side of the straight line passing through the plot (evaluated at $k=\text{\pounds{}}20,000$) denotes the sustainability area. For the CEACs, in addition to the results under MAR (solid lines), the probability values for the four MNAR models described in~\S\ref{MNAR} are represented with different types of dashed lines.}\label{CEAC}
\end{figure}

The CEP (Figure~\ref{CEAC}, panel a) is a graphical representation of the joint distribution for the population average effectiveness and costs increments, indicated respectively as $\Delta_e=(\mu_{e2}-\mu_{e1})$ and $\Delta_c=(\mu_{c2}-\mu_{c1})$, under the three model specifications (light red for the Bivariate Normal, light green for the Beta-Gamma and light blue for the Hurdle Model). The slope of the straight line crossing the plane is the ``willingness to pay'' threshold (often indicated as $k$), and can be considered as the amount of budget the decision-maker is willing to spend to increase the health outcome of one unit and effectively is used to trade clinical benefits for money. Points lying below this straight line fall in the so-called \textit{sustainability area} \citep{Baioa} and suggest that the active intervention is more cost-effective than the control. This is because in this area the new intervention is either more effective and less expensive (in the South-Eastern quadrant) or it produces an increase in benefits that more than offsets the increase in the costs (points in the North-Eastern quadrant below the line). In the graph, which for simplicity only displays the results associated with the all cases under MAR, we also show the Incremental Cost-Effectiveness Ratio (ICER) computed under each model, as darker colour dots. This is defined as 
$$\mbox{ICER}=\frac{\mbox{E}[\Delta_c]}{\mbox{E}[\Delta_e]}$$ 
and it quantifies the cost per incremental unit of QALYs. For all three models more than 70\% of the samples fall in the sustainability area and are associated with negative ICERs, suggesting that the intervention can be considered as cost-effective by producing a QALYs gain at virtually no extra costs, or even saving money.

The CEAC (Figure~\ref{CEAC}, panel b) is obtained by computing the proportion of points lying in the sustainability area upon varying the willingness to pay threshold $k$. Based on general recommended guidelines \citep{NICE2013}, we consider a range for $k$ up to \pounds{30,000} per QALY gained.  The CEAC estimates the probability of cost-effectiveness, thus providing a simple summary of the uncertainty associated with the ``optimal'' decision-making suggested by the ICER. For each model, the results under MAR are reported using solid lines with different colours, i.e.~red for the Bivariate Normal, green for the Beta-Gamma and blue for the Hurdle Model. In addition, the results associated with the four MNAR scenarios are reported using different types of dashed~lines. Under MAR, for the Bivariate Normal and Beta-Gamma models the CEACs indicate the cost-effectiveness of the new intervention with a probability above $0.8$ for most values of $k$. Conversely, under the Hurdle Model, the curve is shifted downward by $0.24$ and $0.16$  with respect to the Bivariate Normal and Beta-Gamma models, respectively, and suggests a more uncertain~conclusion. Perhaps unsurprisingly, none of these results is robust to the alternative MNAR scenarios explored. The CEAC plot clearly shows a large sensitivity of the cost-effectiveness probability with respect to the assumed number of structural ones in both treatment groups. Indeed, the curves span a huge probability range from~$0.2$ under MNAR4 to $1$ under MNAR3. This implies a considerable change in the output of the decision process and severely undermines the validity of the conclusions obtained under~MAR.

\section{Discussion}\label{discussion}
In CEAs alongside RCTs, analysts typically rely on standard models that ignore or at best fail to properly account for potentially important features in the data, such as the correlation between costs and effectiveness, skewness in the distribution of the observed data, the presence of structural values and, almost invariably, missing data. In this paper, we have presented a general Bayesian framework that is able to overcome these problems. 

The analysis of our motivating example shows notable variations in the results, compared with those of the original analysis. Accounting for the structural ones and missingness uncertainty has a considerable impact on the cost-effectiveness of the new intervention and future research prioritisation. Our results are obtained with specific reference to the motivating example. However, the MenSS study is very much representative of the ``typical'' dataset used in CEAs alongside RCTs. Thus, it is highly likely that the same features (and potentially the same contradictions in the results, upon varying the complexity of the modelling assumptions) apply to many real~cases. This is a very important, if somewhat overlooked problem, as it can thwart the validity of simplistic models that, while established among practitioners, may lead to misleading cost-effectiveness conclusions and bias the decision-making~process.

Missing data pose a serious threat to the economic evaluation as, when confronted with a partially-observed dataset, each analysis makes assumptions about the missing values that cannot be verified from the data at hand. Any measure of fit or predictive accuracy, such as the DIC or Posterior Predictive Checks \citep{Gelman2}, can only provide information about the observed data and therefore tell just part of the story \citep{Celeux,Masonb}. Thus, the use of sensitivity analysis to explore the impact on the results of different plausible missingness assumptions, including MNAR, becomes essential. The Bayesian approach naturally allows to perform these assessments through the incorporation of external evidence (e.g.~expert opinions) in the model using prior distributions while ensuring consistency and the correct propagation of uncertainty throughout the model.

We have demonstrated one possible way of assessing the robustness of the results to a range of MNAR scenarios. Even though our approach assumes specific MNAR values (structural ones), it has the advantage of being easy to implement and offers a starting point to investigate MNAR assumptions more thoroughly. Specifically, if one of these scenarios is thought to be more realistic, then it can be explored using more advanced methods that explicitly allow for variability in the MNAR values, e.g.~Selection Models or Pattern Mixture Models~\citep{Daniels,Mason,Molenberghsb}. 

Finally, a potentially relevant question concerning missing variables derived from repeated questionnaires, e.g.~EQ-5D, is whether imputation should be carried out at the time scores (utilities) or total scores (QALYs) level. The two approaches may substantially differ as imputing at the time scores level typically requires that the longitudinal structure of the data is appropriately modelled. Similar issues apply to multi-item questionnaires. The performance of these two alternative imputation strategies has only recently been compared in the health economic literature and further research is needed \citep{Lambert,Eekhout,Simons}. 

In conclusion, in this work we have presented a flexible Bayesian analytic framework that can: \textit{a)} jointly model costs and effectiveness; \textit{b)} account for skewness and structural values; and \textit{c)} assess the robustness of the results under a set of differing plausible missingness assumptions. These are typical features affecting CEA individual-level data that should be simultaneously addressed to avoid biased results, which may in turn lead to misleading cost-effectiveness conclusions. The availability of methodological and practical tools such as the ones presented in this paper have the potential to improve the work of modellers and regulators alike, thus advancing the fields of economic evaluation of health care interventions.

\section*{Acknowledgement}
\ack{Mr Andrea Gabrio is partially funded in his PhD programme at University College London by a research grant sponsored by The Foundation BLANCEFLOR Boncompagni Ludovisi, n\'{e}e Bildt.}

\ack{Dr Gianluca Baio is partially supported as the recipient of an unrestricted research grant sponsored by Mapi Group at University College London.}

\ack{Finally, we wish to thank Ms Julia V. Bailey and Ms Rachael Hunter at University College London for providing the MenSS trial data and advise on the original economic model.}

\bibliographystyle{wileyNJD-AMA}
\bibliography{wileyNJD-AMA}

\begin{thebibliography}{10}

\bibitem{NICE2013}
NICE . {\it Guide to the Methods of Technological Appraisal}.
\newblock London, UK: NICE; 2013.

\bibitem{Briggs2000}
Briggs A. Handling uncertainty in cost-effectiveness models.  {\it
  PharmacoEconomics. }2000;22:479-500.

\bibitem{OHagan2004}
OHagan A, McCabe C, Hakehurst R, et al. Incorporation of uncertainty in health
  economic modelling studies.  {\it PharmacoEconomics. }2004;23:529-536.

\bibitem{Sculpher}
Sculpher M, Claxton K, Drummond M, McCabe C. Whither trial-based economic
  evaluation for health decision making?.  {\it Health Economics.
  }2005;15:677-687.

\bibitem{Spiegelhalter2003}
Spiegelhalter D, Best N. Bayesian approaches to multiple sources of evidence
  and uncertainty in complex cost-effectiveness modelling.  {\it Statistics in
  Medicine. }2003;22:3687-3709.

\bibitem{Claxtonb}
Claxton K. The irrelevance of inference: a decision making approach to
  stochastic evaluation of health care technologies.  {\it Journal of Health
  Economics. }1999;18:342-364.

\bibitem{Baioa}
Baio G. {\it Bayesian Methods in Health Economics}.
\newblock University College London, London, UK: Chapman and Hall/CRC; 2012.

\bibitem{Jackson}
Jackson C, Thompson S, Sharples L. Accounting for uncertainty in health
  economic decision models by using model averaging.  {\it Journal of the Royal
  Statistical Society: Series A. }2009;172:383-404.

\bibitem{Briggs2006}
Briggs A, Schulpher M, Claxton K. {\it Decision Modelling for Health Economic
  Evaluation}.
\newblock Oxford, UK: Oxford university press; 2006.

\bibitem{Spiegelhalterb}
Spiegelhalter DJ, Abrams KR, Myles JP. {\it Bayesian approaches to clinical
  trials and health-care evaluation}.
\newblock John Wiley and Sons; 2004.

\bibitem{Manca}
Manca A, Hawkins N, Sculpher MJ. Estimating mean QALYs in trial-based
  cost-effectiveness analysis: the importance of controlling for baseline
  utility.  {\it Health Economics. }2005;14:487-496.

\bibitem{Hunter}
Hunter RM, Baio G, Butt T, Morris S, Round J, Freemantle N. An Educational
  Review of the Statistical Issues in Analysing Utility Data for Cost-Utility
  Analysis.  {\it PharmacoEconomics. }2015;33:355-366.

\bibitem{Agency}
{European Medicines Agency} . {\it {Committee for Medicinal Products for Human
  Use (CHMP). Guideline on adjustment for baseline covariates}. }
  \url{http://www.ema.europa.eu/docs/en_GB/document_library/Scientific_guideline/2013/06/WC500144946.pdf};
  2013.

\bibitem{OHagan}
O'Hagan A, Stevens JW. A Framework for Cost-Effectiveness Analysis from
  Clinical Trial Data.  {\it Health Economics. }2001;10:303-315.

\bibitem{Nixon}
Nixon RM, Thompson SG. Methods for incorporating covariate adjustment, subgroup
  analysis and between-centre differences into cost-effectiveness evaluations.
  {\it Health Economics. }2005;14:1217-1229.

\bibitem{Thompson}
Thompson SG, Nixon RM. How Sensitive Are Cost-Effectiveness Analyses to Choice
  of Parametric Distributions?.  {\it Medical Decision Making. }2005;4:416-423.

\bibitem{Briggsf}
Briggs A, Gray A. The distribution of health care costs and their statistical
  analysis for economic evaluation.  {\it Health Serv Res Pol. }1998;3:233-245.

\bibitem{Rascati}
Rascati KL, Smith LJ, Neilands T. Dealing with Skewed Data: An Example Using
  Asthma-Related Costs of Medicaid Clients.  {\it Health Economics.
  }2001;23:481-498.

\bibitem{OHagan2003}
O'Hagan A, Stevens JW. Assessing and comparing costs: how robust are the
  bootstrap and methods based on asymptotic normality?.  {\it Health Economics.
  }2003;12:33-49.

\bibitem{Basu}
Basu A, Manca A. Regression Estimators for Generic Health-Related Quality of
  Life and Quality-Adjusted Life Years.  {\it Medical Decision Making.
  }2012;1:56-69.

\bibitem{Cooper}
Cooper N, Sutton AJ, Mugford M, Abrams K. Use of Bayesian Markov Chain Monte
  Carlo Methods to Model Cost-of-Illness Data.  {\it Medical Decision Making.
  }2003;23:38-53.

\bibitem{Mihaylova}
Mihaylova B, Briggs A, O'Hagan A, Thompson SG. Review of Statistical Methods
  for Analysing Healthcare Resources and Costs.  {\it Health Economics.
  }2011;20:897-916.

\bibitem{Ntzoufras}
Ntzoufras I. {\it Bayesian Modelling Using WinBUGS}.
\newblock New York, US: John Wiley and Sons; 2009.

\bibitem{Baio}
Baio G. Bayesian models for cost-effectiveness analysis in the presence of
  structural zero costs.  {\it Statistics in Medicine. }2014;33:1900-1913.

\bibitem{Tooze}
Tooze J, Grunwald G, Jones K. Analysis of repeated measures data with clumping
  at zero.  {\it Statistical Methods in Medical Research. }2002;211:341-355.

\bibitem{Harkanen}
Harkanen T, Maljanen T, Lindfors O, Virtala E, Knekt P. Confounding and Missing
  Data in Cost-Effectiveness Analysis: Comparing different methods.  {\it
  Health Economics Review. }2013;28:3-8.

\bibitem{ISPOR}
Ramsey SD, Willke RJ, Glick H, et al. Cost-Effectiveness Analysis Alongside
  Clinical Trials II-An ISPOR Good Research Practices Task Force Report.  {\it
  Value in Health. }2015;18:161-172.

\bibitem{Groenwold}
Groenwold RHH, Rogier A, Donders T, Roes KCB, Harrell FE, Moons KGM. Dealing
  With Missing Outcome Data in Randomized Trials and Observational Studies.
  {\it American Journal of Epidemiology. }2012;175:210-217.

\bibitem{Wood}
Wood AM, White IR, Thompson SG. Are missing outcome data adequately handled?A
  review of published randomized controlled trials in major medical journals.
  {\it Clinical Trials. }2004;1:368-376.

\bibitem{Noble}
Noble SM, Hollingworth W, Tilling K. Missing data in trial-based
  cost-effectiveness analysis: the current state of play.  {\it Health
  Economics. }2012;21:187-200.

\bibitem{Gabrio}
Gabrio A, Mason AJ, Baio G. Handling Missing Data in Within-Trial
  Cost-Effectiveness Analysis: A Review with Future Recommendations.  {\it
  PharmacoEconomics-Open. }2017;1:79-97.

\bibitem{Briggsb}
Briggs A, Clark T, Wolstenholme J, Clarke P. Missing…. presumed at random:
  cost-analysis of incomplete data.  {\it Health Economics. }2003;12:377-392.

\bibitem{Mancab}
Manca P, Palmer S. Handling Missing Data in Patient-Level Cost-Effectiveness
  Analysis Alongside Randomised Clinical Trials.  {\it Appl Health Econ Health
  Policy. }2005;4:65-75.

\bibitem{Faria}
Faria R, Gomes M, Epstein D, White IR. A Guide to Handling Missing Data in
  Cost-Effectiveness Analysis Conducted Within Randomised Controlled Trials.
  {\it PharmacoEconomics. }2014;32:1157-1170.

\bibitem{Rubina}
Rubin DB. {\it Multiple Imputation for Nonresponse in Surveys}.
\newblock New York, US: John Wiley and Sons; 1987.

\bibitem{DiazOrdaz}
Diaz-Ordaz K, Kenward MG, Grieve R. Handling missing values in cost
  effectiveness analyses that use data from cluster randomized trials.  {\it
  Journal of the Royal Statistical Society: Series A. }2014;177:457-474.

\bibitem{Burton}
Burton A, Billingham LJ, Bryan S. Cost-effectiveness in clinical trials: using
  multiple imputation to deal with incomplete cost data.  {\it Clinical Trials.
  }2007;4:154-161.

\bibitem{Schafer}
Schafer JL. {\it Analysis of Incomplete Multivariate Data}.
\newblock New York, US: Chapman and Hall; 1997.

\bibitem{Schaferbook}
Schafer JL. Multiple imputation for multivariate missing data problems: A data
  analyst's perspective.  {\it Multivariate Behavioural Research.
  }1998;33:545-571.

\bibitem{VanBuuren}
Van~Buuren S, Groothuis-Oudshoorn K. mice: Multivariate Imputation by Chained
  Equations in R.  {\it Journal of Statistical Software. }2011;45:1-67.

\bibitem{Carpenter}
Carpenter JR, Kenward MG, IR~White. Sensitivity analysis after multiple
  imputation under missing at random: a weighting approach.  {\it Statistical
  Methods in Medical Research. }2007;16:259-275.

\bibitem{Molenberghsb}
Molenberghs G, Fitzmaurice G, Kenward MG, Tsiatis A, Verbeke G. {\it Handbook
  of Missing Data Methodology}.
\newblock Boca Raton, FL: Chapman and Hall; 2015.

\bibitem{Daniels}
Daniels MJ, Hogan JW. {\it Missing Data in Longitudinal Studies: Strategies for
  Bayesian Modeling and Sensitivity Analysis}.
\newblock New York, US: Chapman and Hall; 2008.

\bibitem{Mason}
Mason A, Richardson S, Plewis I, Best N. Strategy for Modelling Nonrandom
  Missing Data Mechanisms in Observational Studies Using Bayesian Methods.
  {\it Journal of Official Statistics. }2012;28:279-302.

\bibitem{Hunterb}
Bailey JV, Webster R, Hunter R, et al. The Mens's Safer Sex project:
  intervention development and feasibility randomised controlled trial of an
  interactive digital intervention to increase condom use in men.  {\it Health
  Technology Assessment. }2016;20.

\bibitem{Drummond}
Drummond MF, Schulpher MJ, Claxton K, Stoddart GL, Torrance GW. {\it Methods
  for the economic evaluation of health care programmes. 3rd ed}.
\newblock Oxford, UK: Oxford University Press; 2005.

\bibitem{Gelmanp}
Gelman A. Prior distributions for variance parameters in hierarchical models.
  {\it Bayesian Analysis. }2006;1:515-533.

\bibitem{BCEABook}
Baio G, Berardi A, Heath A. {\it Bayesian Cost-Effectiveness Analysis with the
  R package BCEA}.
\newblock Springer; 2017.

\bibitem{Plummer}
Plummer M. {\it {JAGS: Just Another Gibbs Sampler}. }
  \url{http://www-fis.iarc.fr/~martyn/software/jags/}; 2010.

\bibitem{Su}
YS~Su., Yajima M. {\it {Package ‘R2jags’}. }
  \url{http://www-fis.iarc.fr/~martyn/software/jags/}; 2015.

\bibitem{Gelman2}
Gelman A, Carlin J, Stern H, Rubin D. {\it Bayesian Data Analysis - 2nd
  edition}.
\newblock New York, NY: Chapman and Hall; 2004.

\bibitem{Spiegelhalter}
Spiegelhalter DJ, Best NG, Carlin BP, Linde A. Bayesian measures of model
  complexity and fit.  {\it Journal of the Royal Statistical Society.
  }2002;64:583-639.

\bibitem{Celeux}
Celeux G, Forbes S, Robert CP, Titterington DM. Deviance Information Criteria
  for Missing Data Models.  {\it Bayesian Analysis. }2006;1:651-674.

\bibitem{Masonb}
Mason A, Richardson S, Best N. Two-pronged Strategy for Using DIC to Compare
  Selection Models with Non-Ignorable Missing Responses.  {\it Bayesian
  Analysis. }2012;7:109-146.

\bibitem{Black}
Black WC. A Graphic Representation of Cost-Effectiveness.  {\it Medical
  Decision Making. }1990;10:212-214.

\bibitem{VanHout}
Van~Hout BA, Al~MJ, Gordon GS, Rutten FFH, Kuntz KM. Costs, Effects and
  C/E-Ratios Alongside a Clinical Trial.  {\it Health Economics.
  }1994;3:309-319.

\bibitem{Lambert}
Lambert PC, Billingham LJ, Cooper NJ, Sutton AJ, Abrams KR. Estimating the
  cost-effectiveness of an intervention in a clinical trial when partial cost
  information is available: a Bayesian approach.  {\it Health Economics.
  }2008;17:67-81.

\bibitem{Eekhout}
Eekhout I. {\it Don't Miss Out!: Incomplete data can contain valuable
  information}.
\newblock Amsterdam, NL: EMGO+ Institute for Health and Care Research,
  Department of Epidemiology and Biostatistics, VU University Medical Center;
  2014.

\bibitem{Simons}
Simons CL, Rivero-Arias O, Yu~LM, Simon J. Multiple imputation to deal with
  missing EQ-5D-3L data: Should we impute individual domains or the actual
  index?.  {\it Qual Life Res. }2015;24:805-815.

\end{thebibliography}

\clearpage

\begin{appendices}
\section{Model Code and Implementation}\label{code}

\subsection{Implementation ``trick''}
The model described in~\S\ref{HU} uses a different sampling distribution for the QALYs, depending on the observed value of the indicator $d_{ie}$
\begin{equation*}
e_{i}\mid d_{ie} \sim \begin{cases}
      p(e_{i}\mid d_{ie}=0)=p(e_{i}\mid \bm{\theta}^{<1}), & \text{if $e_{i}<1$}\\
    p(e_{i}\mid d_{ie}=1)=p(e_{i}\mid \bm{\theta}^{1}), & \text{if $e_{i}=1$},
  \end{cases}
\end{equation*}
where the model for $e_i=1$ is degenerate at a point mass at 1, while that for $e_i<1$ is defined in terms of a Beta distribution. We can conveniently re-write this more succinctly and with specific reference to our case as 
\[e_i \sim \mbox{Beta}\left(\phi_{ie}^{d_{ie}}\tau_{ie}^{d_{ie}}, \left(1-\phi_{ie}^{d_{ie}}\right)\tau_{ie}^{d_{ie}}\right).\]
If we set $\phi_{ie}^{1}=1$ and select $\tau_{ie}^{1}$ in order to induce a variance as close to 0 as possible, the two specifications are identical. Unfortunately, it is not possible to do so in the \texttt{BUGS}/\texttt{JAGS} language, because the Beta distribution is specified in the open interval $(0,1)$ and thus setting $\phi^1_{ie}=1$ implies that $\tau^1_{ie}=0$, which is not allowed. 

However, the required behaviour is very closely mimicked if we define our model~with
\[ \mbox{logit}(\phi_{ie}^{1})=\alpha^1_0\,[+\ldots] \]
and set $\alpha^1_0=\mbox{logit}(0.999999)$ and $\sigma_e\approx 0$, which implies $\mu_e\approx 1$ with virtually no uncertainty. In other words, we can specify extremely informative priors on the parameters $\bm\theta^1$ so that the implied distribution for the structural ones components of the mixture is concentrated around 1 with essentially no uncertainty. More importantly, with such a prior no amount of data can modify the posterior. The critical aspect of this strategy, however, is that inferences may be potentially sensitive to the way such priors are specified, that is whether a small variation in the hyperprior values can affect the posterior estimates.

In fact, the estimation of the other parameters is not really affected by this choice, provided that the encoded prior really induces the variance towards zero. It is also plausible that different values for $\sigma^{1}_{e}$ have an impact on measures of model fit, such as the DIC. This is essentially due to the fact that the population is really comprised of two groups, one of which shows QALYs that are identically one. Thus, the closer the approximation to zero for the variance the better the fit to the observed data and therefore the smaller the resulting~DIC.

With this in mind, we have used different values for $\sigma^{1}_{e}$ to assess the impact on the mean QALYs estimates. Fixing the value of the mean for the ones group to $\mu^{1}_{e}=0.999999$ corresponds to an upper bound for the standard deviation of $0.0001$ (see~\S\ref{BG}). We have explored a range of possibilities by progressively decreasing this value and assessed their impact on posterior results. 

\begin{figure}[!h]
\centering
\subfloat{\includegraphics[scale=1.5]{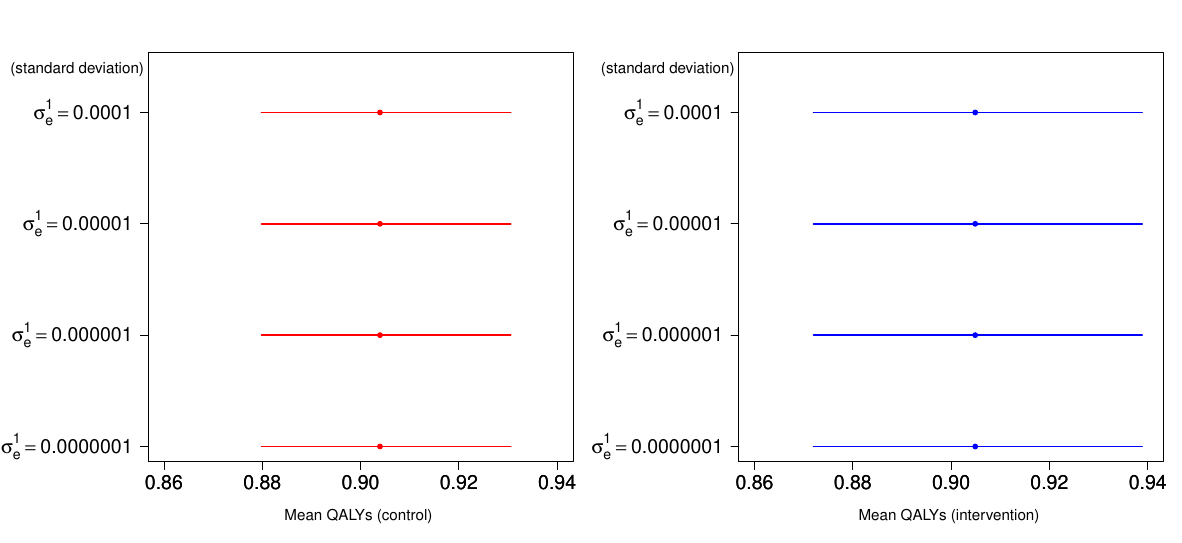}}
\caption{Sensitivity analysis for the choice of the standard deviation for the distribution of the structural ones in the QALYs. For each value of $\sigma^{1}_{e}$ tested, posterior means and 90\% HPD intervals for the mean QALYs parameters are respectively represented with dots and lines (red for the control and blue for the intervention group).}\label{SA}
\end{figure}

Figure~\ref{SA} shows the sensitivity of the inferences across the alternative specifications for $\sigma^{1}_{e}$. Results in terms of mean posterior estimates and 90\% HPD intervals were almost unchanged in all the cases. Thus, we can assert that model performance was unaffected by the choice of the value for $\sigma^{1}_{e}$. We also observe that the DIC becomes smaller when the standard deviation parameter decreases and the best-fitting model is the one associated with the smallest values, although the results are hardly different from both an estimation and convergence perspective for all the parameters. 

\begin{center}
FIGURE 7 HERE
\end{center}

\subsection{Code}
The complete \texttt{JAGS} code for the Hurdle Model used in the analysis is given below.
\begin{Verbatim}[tabsize=8,fontsize=\footnotesize]
model {

# data variables
# e, c and u denote the QALYs, costs and baseline utilities
# d.e and d.u denote the structural one indicators for e and u
# age, ethnicity and employment are covariates in the model of d.e and d.u

# control group (t = 1)

	for(i in 1 : N1) {
	
		# 1. Module for the structural ones in the QALYs 
		d.e1[i] ~ dbern(pi.e[i, 1])
		logit(pi.e[i, 1]) <- gamma0[1] + gamma1[1] * (u1[i] - mean(u1[])) + 
			gamma2[1] * (age1[i] - mean(age1[])) + gamma3[ethnicity1[i], 1] + gamma4[employment1[i], 1]

		#2. Module for the structural ones in the baseline utilities 
		d.u1[i] ~ dbern(pi.u[i, 1])
		logit(pi.u[i, 1]) <- eta0[1] + eta1[1] * (age1[i] - mean(age1[])) + eta2[ethnicity1[i], 1] + eta3[employment1[i], 1]

		#3. Marginal module for the QALYs
		e1[i] ~ dbeta(phi.e[i, 1] * tau.e[i, 1], (1 - phi.e[i, 1]) * tau.e[i, 1])
		tau.e[i, 1] <- phi.e[i, 1] * (1 - phi.e[i, 1]) / pow(sigma.e[d.e1[i] + 1], 2) - 1
		logit(phi.e[i, 1]) <- alpha0[d.e1[i]+1, 1] + alpha1[d.e1[i]+1, 1] * (u1[i] - mean(u1[]))

		#4. Marginal module for the baseline utilities
		u1[i] ~ dbeta(mu.u[d.u1[i] + 1, 1] * tau.u[d.u1[i] + 1, 1], (1 - mu.u[d.u1[i] + 1, 1]) * tau.u[d.u1[i] + 1, 1]) 
      
		#5. Conditional module for the costs
		c1[i] ~ dgamma(phi.c[i, 1] * tau.c[i, 1], tau.c[i, 1])
		tau.c[i, 1] <- phi.c[i, 1] / pow(sigma.c[1], 2)
		log( phi.c[i, 1]) <- beta0[1] + beta1[1] * (e1[i] - mu.e[1])
	}

#intervention group (t = 2)

	for(i in 1 : N2) {
	
		#1. Module for the structural ones in the QALYs 
		d.e2[i] ~ dbern(pi.e[i, 2])
		logit(pi.e[i, 2]) <- gamma0[2] + gamma1[2] * (u2[i] - mean(u2[])) + 
			gamma2[2] * (age2[i] - mean(age2[])) + gamma3[ethnicity2[i], 2] + gamma4[employment2[i], 2]

		#2. Module for the structural ones in the baseline utilities 
		d.u2[i] ~ dbern(pi.u[i, 2])
		logit(pi.u[i, 2]) <- eta0[2] + eta1[2] * (age2[i] - mean(age2[])) + eta2[ethnicity2[i], 2] + eta3[employment2[i], 2]

		#3. Marginal module for the QALYs
		e2[i] ~ dbeta(phi.e[i, 2] * tau.e[i, 2], (1 - phi.e[i, 2]) * tau.e[i, 2])
		tau.e[i,2] <- phi.e[i, 2] * (1 - phi.e[i, 2]) / pow(sigma.e[d.e2[i] + 1], 2) - 1
		logit(phi.e[i, 2]) <- alpha0[d.e2[i] + 1, 2] + alpha1[d.e2[i] + 1, 2] * (u2[i] - mean(u2[]))

		#4. Marginal module for the baseline utilities
		u2[i] ~ dbeta(mu.u[d.u2[i] + 1, 2] * tau.u[d.u2[i] + 1, 2], (1 - mu.u[d.u2[i] + 1, 2]) * tau.u[d.u2[i] + 1, 2]) 
      
		#5. Conditional module for the costs
		c2[i] ~ dgamma(phi.c[i, 2] * tau.c[i, 2], tau.c[i, 2])
		tau.c[i, 1] <- phi.c[i, 1] / pow(sigma.c[2], 2)
		log( phi.c[i, 2]) <- beta0[2] + beta1[2] * (e2[i] - mu.e[2])
	}
      
#Priors
#priors for module 1 and 2

	for(t in 1 : 2) {
		gamma0[t] ~ dlogis(0, 1)
		gamma1[t] ~ dnorm(0, 0.00001)
		gamma2[t] ~ dnorm(0, 0.00001)
      
		eta0[t] ~ dlogis(0, 1)
		eta2[t] ~ dnorm(0, 0.00001)
      
		#priors on coefficients for categorical covariates 
		#(setting reference category as 0)
		gamma3[1, t] <- 0
		gamma4[1, t] <- 0
      
		eta2[1, t] <- 0
		eta3[1, t] <- 0
	}
      
# set priors for all other categories
# use blocking to improve model convergence
# mu and tau values provided as data variables with zero means and small precisions (0.00001)
# ethnicity has different numbers of categories between arms

	gamma3[2:14, 1] ~ dmnorm(mu1.gamma3[], tau1.gamma3[, ])
	gamma3[2:12, 2] ~ dmnorm(mu2.gamma3[], tau2.gamma3[, ])
	gamma4[2:6, 1] ~ dmnorm(mu1.gamma4[], tau1.gamma4[, ])
	gamma4[2:6, 2] ~ dmnorm(mu2.gamma4[], tau2.gamma4[, ])
      
	eta2[2:14, 1] ~ dmnorm(mu1.eta2[], tau1.eta2[, ])
	eta2[2:12, 2] ~ dmnorm(mu2.eta2[], tau2.eta2[, ])
	eta3[2:6, 1] ~ dmnorm(mu1.eta3[], tau1.eta3[, ])
	eta3[2:6, 2] ~ dmnorm(mu2.eta3[], tau2.eta3[, ])

	for(t in 1 : 2) {
		# priors for model 3
		# priors for the ones group in the QALYs
		alpha0[2, t] <- logit(0.999999)
		alpha1[2, t] <- 0
		sigma.e[2, t] <- 0.00001   
		# priors for the non-ones group in the QALYs 
		alpha0[1, t] ~ dnorm(0, 0.000001)
		alpha1[1, t] ~ dnorm(0, 0.000001)
		sigma.e[1, t] ~ dunif(0, sd.limit.e[t])
		sd.limit.e[t] <- pow(mu.e[1, t] * (1 - mu.e[1, t]), 0.5)

		# priors for model 4
		# priors for the ones group in the baseline utilities
		tau.u[2, t] <- mu.u[2, t] * (1 - mu.u[2, t]) / pow(sigma.u[2, t], 2) - 1
		logit(mu.u[2, t]) <- delta0[2, t]
		delta0[2, t] <- logit(0.999999)
		sigma.u[2, t] <- 0.00001    
		# priors for the non-ones group in the baseline utilities     
		tau.u[1, t] <- mu.u[1,t] * (1 - mu.u[1, t]) / pow(sigma.u[1, t], 2) - 1
		logit(mu.u[1, t]) <- delta0[1,t]
		delta0[1, t] ~ dnorm(0, 0.00001)
		sigma.u[1, t] ~ dunif(0, sd.limit.u[t])
		sd.limit.u[t] <- pow(mu.u[1, t] * (1 - mu.u[1, t]), 0.5)
      
		# priors for module 5
		beta0[t] ~ dnorm(0, 0.00001)
		sigma.c[t] ~ dunif(0, 1000)
		beta1[t] ~ dnorm(0, 0.00001)

		# obtain marginal probabilities for weighting
		p[t] <- ilogit(gamma0[t])
            
		# obtain the weighted marginal mean QALYs
		mu.e[t] <- p[t] + (1-p[t]) * ilogit(alpha0[t])
	}

	# compute incremental QALYs and costs
	Delta_e <- mu.e[2] - mu.e[1]
	Delta_c <- mu.c[2] - mu.c[1]
}
\end{Verbatim}

\end{appendices}

\clearpage

\begin{table}[!h]
\centering
\begin{tabular}{cccc}
\toprule
\textbf{Time} & \textbf{Type of outcome} & \multicolumn{1}{c}{\textbf{Control} ($n_{1}$=75)} & \multicolumn{1}{c}{\textbf{Intervention} ($n_{2}$=84)}\\  \midrule
 & & observed (\%)  &  observed (\%)\\ [0.5em]
 Baseline & utilities & 72 (96\%)  & 72 (86\%) \\ 
 3 months & utilities and costs & 34 (45\%) & 23 (27\%) \\ 
 6 months  & utilities and costs & 35 (47\%) & 23 (27\%)\\ 
 12 months  & utilities and costs & 43 (57\%) & 36 (43\%) \\ \midrule
\bf{complete cases}& utilities and costs & 27 (44\%) & 19 (23\%) \\
\bottomrule
\end{tabular}
\caption{Number and proportion of observed cases at each time point for the utility and cost data (self-recorded questionnaires), presented by trial group (baseline data only related to the utilities). The number of individuals having valid data at each time point (complete cases) is also reported at the bottom of the table. Over the trial period both drop-out and intermittent missingness occur; at each time point only unit-nonresponse patterns are~observed.}\label{Tpattern}
\end{table}

\begin{table}[!h]
\centering
\begin{tabular}{c|c|c}
\toprule
\textbf{Scenario} & \textbf{Control}  ($n^{*}_{1}=13$) &  \textbf{Intervention} ($n^{*}_{2}=22$) \\ \midrule
\textbf{MNAR1} & $d_{ie}=1$ & $d_{ie}=1$ \\ [0.5em]
\textbf{MNAR2}  & $d_{ie}=0$  & $d_{ie}=0$ \\ [0.5em]
\textbf{MNAR3} & $d_{ie}=1$  & $d_{ie}=0$ \\ [0.5em]
\textbf{MNAR4} & $d_{ie}=0$ & $d_{ie}=1$ \\ 
\bottomrule
\end{tabular}
\caption{Alternative MNAR scenarios considered in the MenSS study for the Hurdle Model. In each scenario, individuals who are potentially associated with a unit QALYs in the control ($n^{*}_{1}=13$) and intervention ($n^{*}_{2}=22$) group are assigned to either the structural or non-structural components by setting the value of the indicator $d_{ie}$ equal to 1 or 0, respectively.}\label{tmnar}
\end{table}

\clearpage

\end{document}